\documentstyle[eqsecnum,aps,epsf]{revtex} 

\begin{document}
\twocolumn[\hsize\textwidth\columnwidth\hsize\csname@twocolumnfalse\endcsname

\title{Positronium scattering by atoms and molecules at low energies} 

\author{Sadhan K. Adhikari\\
Instituto de F\'{\i}sica Te\'orica, 
Universidade Estadual Paulista, \\
01.405-900 S\~ao Paulo, S\~ao Paulo, Brazil}

\date{\today}
\maketitle

\begin{abstract}

The recent theoretical and experimental activities in positronium (Ps)
scattering by atoms and molecules are reviewed with special emphasis at
low energies. We critically compare the results of different groups $-$
theoretical and experimental. The theoretical approaches considered
include  the $R$-matrix and close-coupling methods applied to Ps-H,
Ps-He, and Ps-Li scattering, and  a coupled-channel approach with a
nonlocal model-potential for Ps scattering by H, He, H$_2$, Ne, Ar, Li,
Na, K, Rb, Cs, and Ps and for pickoff quenching in Ps-He scattering.  Results
for scattering lengths, partial, total and differential cross sections as
well as resonance and binding energies in different systems are discussed.


\end{abstract}

\vskip1.5pc]

\section{Introduction}

Due to the technical advancement in the preparation of the ortho
positronium (Ps) beam, precise experimental results for the total cross
section
of Ps scattering by different atomic and molecular targets are now
available \cite{1,2a,2b,3,4}. Among recent experiments there have been
measurements of total cross section by Laricchia's group for energies
up to 100 eV for H$_2$, He, and Ar \cite{1,2a,2b}.  Gidley's group
provided results for low-energy ($\sim $ 1 eV) elastic cross section for
Ps scattering by H$_2$, N$_2$, He, Ar, Ne, isobutane, and neopentane
\cite{3}.  They also studied the collisional quenching rate of ortho Ps at
different temperatures for these targets as well as for ethane and methane
\cite{4ab}. Nagashima et al. provided cross section for
Ps-He scattering at 0.15 eV \cite{4}. Hyodo provided  new results
for quenching rate on different targets at this workshop \cite{hy}.
Among the older experiments there
are results for zero-energy Ps-He cross section  and pickoff quenching
\cite{4a,11a,11b,11c} as well as low-energy cross-section of Ps scattering
by
noble
gases \cite{4aa}.
There are several comprehensive reviews on this subject
\cite{5a,5b,5c,5d}.

On the theoretical front, after the pioneering study of Ps-H scattering by
Massey and Mohr in 1954 \cite{mm} using the first Born approximation with
Oppenheimer exchange \cite{bo}, there have been studies of Ps-H and
Ps-He scattering using the static exchange approximation, respectively, by
Fraser
\cite{14a,15a,14,15}, and by Bransden \cite{12,13} and their collaborators
in the decades of 1960 and 1970.  Drachman and Houston performed model
calculations of Ps-H \cite{dh,dh2} and Ps-He \cite{16} scattering in the
decade of 1970. There are also accurate calculations  of resonance
\cite{ho5,ho6,ho7,ho3,dra}
and binding energies \cite{be1,be2,be4,be5,be6,be8,be10,10c,10a}
of
Ps-H. 

Extensive
theoretical efforts on Ps-H and Ps-He scattering started in the decade of
1990 with
the coupled channel $R$-matrix and close-coupling (CC) approaches by
Walters
\cite{5c,6,7} and Ghosh \cite{8a,8b,8c,8d,9,10,9a,9b,h1,h2} and their
collaborators.  More recently, there has been successful calculation of Ps
scattering by H, He, Ne, Ar, Li, Na, K, Cs, Rb, H$_2$ and Ps using a model
exchange potential in a coupled-channel formalism by Biswas and this
author \cite{17a,17b,18,19,20,21,21a,21b}. This latter study produced,
in addition to cross sections, results for resonance and binding energies
of different Ps-atom systems \cite{pkb,ska3,ska4} as well as for pickoff
quenching rate of the interaction of ortho Ps with He \cite{22}.

Of the different Ps-atom systems, Ps-He is the most studied system both
theoretically and experimentally and hence deserves special attention in
addition to the most fundamental Ps-H system on which there are some
accurate results for PsH binding
\cite{be1,be2,be4,be5,be6,be8,be10,10c,10a} and resonance energies
\cite{ho5,ho6,ho7,ho3}. However, 
there is considerable discrepancy among the different theoretical
Ps-He cross sections at zero energy.

On the experimental front, there have been conflicting results for the
low-energy Ps-He elastic cross section by Nagashima {\it et al.} \cite{4},
who measured a cross section of $(13\pm 4)\pi a_0^ 2$ at 0.15 eV, by
Coleman {\it et al.} \cite{4aa}, who reported $9\pi a_0^ 2$ at 0 eV, by
Canter {\it et al.} \cite{4a}, who found $8.47\pi a_0^ 2$ at 0 eV, and by
Skalsey {\it et al.} \cite{3}, who measured $(2.6\pm 0.5)\pi a_0^ 2$ at
0.9 eV.  It is unlikely that all these findings could be consistent with
each
other.

The results for the total cross section of Ps-He scattering obtained from
the
coupled-channel calculation employing the model exchange potential
\cite{18,19,20}
are in agreement with experiments of Refs. \cite{1,2a,2b,3} at low
energies
as well as with a variational scattering length \cite{ska2}.
This model, while agrees \cite{18,19,20} with the experimental
total cross sections \cite{1,3} in the energy range 0 to 70 eV, reproduces
\cite{22} successfully the experimental pickoff quenching rate
\cite{4a,11a,11b,11c}.  All other calculations could not reproduce the
general
trend of cross sections of Ps-He scattering in this energy range 
and yielded a much too small quenching rate at low energies
\cite{12,14,22}.  However, the very low-energy elastic cross sections of
the model-potential calculation \cite{18,19,20} are at variance with the
experiments of Refs. \cite{4,4a,4aa}.

In the Ps-H system there are no experimental results of
scattering. However, there are theoretical 
calculations on Ps-H binding \cite{be1,be2,be4,be5,be6,be8,be10}
and resonance 
\cite{ho5,ho6,ho7,ho3} energies
and scattering
lengths \cite{ska1}. These should be considered as guidelines for
testing the coupled-channel calculations using the 
$R$-matrix, CC, and model-potential approaches.
Of these only the model-potential calculation could provide nearly
converged results for the binding and resonance energies of PsH.

Of the three coupled-channel methods only the model-potential approach has
seen further successful applications in Ps scattering by Ne, Ar \cite{20},
Li \cite{pkb}, Na,
K \cite{ska3}, Rb, Cs \cite{ska4}, H$_2$ \cite{21,21a,21b}, and Ps
\cite{mp1}. There has been proposals \cite{mls,go} of Bose-Einstein
condensation using
spin-polarized ortho Ps atoms and a prior knowledge of Ps-Ps scattering
length is of advantage.  For Ne and Ar
the
low-energy results for elastic
cross section are in agreement with recent experiment \cite{20}, whereas
for the
alkali-metal atoms new low-energy resonances have been predicted.  For Ps
scattering by H$_2$, good agreement with experimental total cross section
has been obtained at low to medium energies  \cite{21,21a}.

\section{Model-Potential Approach}

The theory for the study of Ps scattering by $R$-matrix
\cite{5c}
and CC  \cite{8a,8b} approaches  has appeared in the literature and we
refer
the interested readers to appropriate places. Here we only outline the
model-potential approach for Ps-H scattering \cite{17b}. The modifications
for more
complex targets are straightforward and can be found elsewhere
\cite{19,20}. 
The following Lippmann-Schwinger scattering integral equation is
considered in momentum
space
\begin{eqnarray}\label{zzz}
f^\pm _{\mu '\nu ', \mu \nu }({\bf k', k})= B^\pm _{\mu '\nu ', \mu \nu
}({\bf k', k})\nonumber \\
- \sum_{\mu '', \nu ''} \int \frac{d\bf k ''}{2\pi ^2}\frac
{B^\pm _{\mu '\nu ', \mu ''\nu'' }({\bf k', k''}) f^\pm _{\mu ''\nu '',
\mu 
\nu }(\bf k'', k)}{k^2_{\mu ''\nu ''}/4 - k ''^2/4+i0},
\end{eqnarray}
where the singlet (+) and triplet ($-$) ``Born'' amplitudes $B^\pm$ are
given
by $ B^\pm _{\mu '\nu ', \mu \nu
}({\bf k', k})=  B^D _{\mu '\nu ', \mu \nu
}({\bf k', k})\pm  B^E _{\mu '\nu ', \mu \nu
}({\bf k', k})$, where $ B^D$ and $ B^E$ represent the direct and exchange
Born amplitudes and $f^\pm$ are the singlet and the triplet scattering
amplitudes, respectively. The quantum states $\mu$ and $\nu$ refer to the
hydrogen and Ps atoms, respectively,    ${\bf k}, {\bf k'}$
etc. are the appropriate momentum variables; ${ \bf k_{\mu ''\nu ''}}$ is
the on-shell Ps-H relative momentum  in the channel $\mu '' \nu ''$. We
use atomic units  (a.u.) $\hbar=m=e=1$ where $m$ is the electron mass and
$e$ its
charge. The differential cross section for scattering from ${\bf k} \mu
\nu \to {\bf 
k}' \mu '\nu '$ is given by
\begin{equation}
\left(\frac{d\sigma}{d\Omega}  \right)_{\mu '\nu ', \mu
\nu}=\frac{k'}{4k}
[|f^+ _{\mu '\nu ', \mu \nu }({\bf k', k})           |^2+ 3|f^- _{\mu '\nu ',
\mu \nu }({\bf k', k}) | ^2
 ].\end{equation}
The Ps-H direct Born amplitude is given by
\begin{eqnarray}
B^D _{\mu '\nu ', \mu \nu
}({\bf k', k}) &=& \frac{4}{Q^2}\int\phi^* _{\mu '}({\bf r})[1-\exp(i{\bf
Q.r})]
\phi _{\mu }({\bf r})d{\bf r}\nonumber \\
& \times &\int \chi^* _{\nu '}({\bf t})2i\sin({\bf Q.t}/2)\chi_{\nu }({\bf
t})d{\bf t}. 
\end{eqnarray}
The Ps-H model exchange (Born) amplitude is a generalization of the
electron-hydrogen model exchange potential of Ochkur \cite{oc} and Rudge
\cite{ru}
and is given by
\begin{eqnarray}
B^D _{\mu '\nu ', \mu \nu
}({\bf k', k}) &=& \frac{4(-1)^{l +l'}}{D}\int\phi^* _{\mu' }({\bf
r})\exp(i{\bf Q.r})
\phi _{\mu }({\bf r})d{\bf r}
\nonumber \\
&\times &\int \chi^* _{\nu ' }({\bf t})\exp(i{\bf Q.t}/2)\chi_{\nu }({\bf
t})d{\bf t}
\end{eqnarray}
with 
\begin{eqnarray}\label{p}
D=\frac{k^2+k'^2}{8}+ C^2\left[\frac{\alpha_\mu ^2+ \alpha_{\mu'} ^2}{2}+
\frac{\beta_\nu ^2+
\beta_{\nu '}^2}{2}\right],
\end{eqnarray}
where $l$ and $l'$ are the angular momenta of the initial and final Ps
states;
${\bf Q =k - k' }$; $\alpha_\mu ^2/2$, $\alpha_{\mu'} ^2/2$, $\beta_\nu
^2$,
and $\beta_{\nu'}  ^2$ are the binding energies of the initial and final
states
of H and Ps in a.u., respectively; and $C$ is the only parameter of the
potential. Normally, this parameter is to be taken to be unity, which
leads to reasonably good results. However, it can be varied slightly from
unity to get a precise fit to a low-energy observable. 
After  a partial-wave projection the coupled-channel scattering equations
(\ref{zzz}) are solved by the method of matrix inversion. 

\section{Numerical Results}

\subsection{Ps-H System}

There are no experimental results for this system. Ps-H scattering has
been studied with the static-exchange model by Fraser \cite{14a,15a}, with
the pseudo-state $R$-matrix approach by McAlinden et al \cite{7}, with the
CC approach by Ghosh and collaborators \cite{8a,8b,8c,8d}, using a
model-potential by Drachman and Houston \cite{dh,dh2}, and finally by the
coupled-channel model potential approach by Biswas and this author
\cite{17a,17b}. In addition to these scattering calculations there exist
variational calculations for PsH binding energy
\cite{be1,be2,be4,be5,be6,be8,be10} and Ps-H scattering lengths
\cite{ska1,mt}. There are calculations for PsH resonance energies and
width
in
different partial waves at low energies using the complex-coordinate
rotation method \cite{ho5,ho6,ho7,ho3,dra}.

The most accurate  
result  for PsH binding (= 1.064661  eV) seems to be  due
to Frolov and Smith,
Jr. \cite{be8}.
Relativistic effects on PsH has also been studied \cite{10a} where a
binding energy of  1.06404168 eV has been predicted.
Schrader et al. confirmed the existence of PsH experimentally \cite{10c}.
The S-wave resonances in the Ps-H  system were studied by Drachman 
\cite{dra}
and by Yan and Ho \cite{ho5,be6}. The most accurate results for 2S and
3S energies (width) 
are $4.0058\pm 0.0005$   eV ($0.0952\pm 0.0011$ eV) and 
$4.9479\pm 0.0014$   eV ($0.0585\pm 0.0027$ eV) 
\cite{ho5}, respectively. The P- and D-wave
resonances for this system were also studied by  Drachman \cite{dra}  and
by Yan  and Ho \cite{ho6,ho7}. The most accurate values for the 2P and 3P
resonance energies
(widths) are $4.2850\pm .0014$ eV ($0.0435\pm 0.0027$ eV)  and $5.0540\pm
0.0027$
eV ($0.0585\pm 0.0054$ eV) \cite{ho6}, respectively; that for the 3D wave
is $4.710\pm 0.0027$
 eV ($0.0925\pm 0.0054$ eV) \cite{ho7}. Ho and Yan \cite{ho3} also
studied
resonances in
higher partial waves.

In the absence of experiments on Ps-H scattering, the above accurate
results could be used as critical tests for the different coupled-channel
calculations.  Campbell et al. \cite{7} performed a 22-pseudo-Ps-state
$R$-matrix calculation and predicted the following resonance energies
(widths) in S, P and D waves, respectively: 4.55 eV (0.084 eV), 4.88 eV
(0.058 eV), and 5.28 eV (0.47 eV).  Compared to the resonances of Ho and
Yan \cite{ho5,ho6,ho7} the agreement is
fair.
The PsH binding energy of Campbell et al. \cite{7} (0.634 eV) 
show
similar convergence when compared with the accurate results 1.064661 eV
\cite{be8}. Their singlet and triplet scattering lengths 
$a_s=5.20a_0$ and $a_t=2.45a_0$ is only in qualitative agreement with
variational
results: 
 $a_s=(3.49\pm 0,20)a_0$
\cite{ska1} and $4.3a_0$ \cite{mt} and  $a_t=(2.46\pm 0,10)a_0$
\cite{ska1} and $2.2a_0$ \cite{mt},
respectively.  Walters et al. presented new results for Ps-H scattering at
this workshop in a $(14\times 14)$ channel model containing 14 Ps states
and 14 H states \cite{wt}. This preliminary calculation shows that their
elastic cross sections are substantially reduced and the scattering
lengths are in better agreement with variational results.

The CC calculations by Ghosh and collaborators \cite{8a,8b,8d} also 
exhibit slow
 convergence.  No resonance or bound states  have been
reported by them.   This makes the
critical comparison difficult.  
Their singlet scattering lengths calculated with different basis states
are always greater than 5.20$a_0$ of Campbell et al. \cite{7} which
demonstrates only fair agreement  with the variational results 
\cite{ska1,mt}. However, they have recently emphasized the importance of
including the hydrogen states in a coupled-channel calculation \cite{8d}.

With an appropriate choice of the parameter $C (=0.784)$ in Eq. (\ref{p})
using a five-state coupled-channel model Biswas and this author  obtained
4.01 eV, 1.067 eV,
3.72$a_0$ for the singlet Ps-H resonance and binding energies and
scattering length \cite{17a,17b}, in better agreement with the
corresponding
variational results: 4.0058 eV \cite{ho5}, 1.064661 eV \cite{be8}, and
3.49$a_0$ \cite{ska1}, respectively. The agreement with the variational
singlet scattering length of Ref. \cite{mt} is only fair.
They fould that for obtaining good convergence of resonance and binding
energies the inclusion of couple of hydrogen states in the expansion
scheme was essential.  They also calculated total, partial and
differential cross sections at different energies \cite{17a,17b}.
The  P-wave resonance of this
model
\cite{17b} at 
5.08 eV agrees poorly with the calculation of Yan and Ho
\cite{ho6}. However,  
this shows a correlated behavior
among the low-energy observables in the singlet S wave, which is expected
for an attractive effective Ps-H interaction of short-range. For an
approximately fixed range of the effective Ps-H interaction, using the
classic idea of the effective-range expansion, the low-energy Ps-H problem
is expected to be determined by a single parameter -- the strength of
interaction. This means that once this parameter is adjusted to fit an
observable of the low-energy S-wave singlet Ps-H system, a satisfactory
description of other low-energy observables follows.  
As in the Ps-H system there are no experimental
results, we next consider the problems of Ps-He and Ps-H$_2$ scattering
where the above  wisdom is turned to good advantage by fitting a
low-energy data, e.g., the 
scattering cross section. Consequently, the model presents a faithful
representation of Ps-He and Ps-H$_2$ scattering at low and
medium energies.

\subsection{Ps-He System}

There are no known resonance and bound state in the Ps-He system. The
variational results for the scattering length are  $(1.0 \pm
0.1)a_0$
and 1.61$a_0$
corresponding to  zero-energy cross sections of (4.0 $\pm$ 0.8)$\pi
a_0^2$ \cite{ska2} and 10.4$\pi a_0^2$ \cite{mt}, respectively.

The static-exchange model by Sarkar and Ghosh \cite{9}, and by
Blackwood {\it et al.} \cite{6} yielded $14.38\pi a_0^2$ (at 0.068 eV),
and $14.58\pi a_0^2$ (at 0 eV), respectively, for the zero-energy 
cross section. The inclusion of more states of Ps in the CC \cite{10} and
$R$-matrix \cite{6} calculations does not change these results
substantially. The static-exchange calculations by Barker and
Bransden \cite{12} yielded 13.04 $\pi a_0^2$ and by Fraser \cite{13}
yielded $14.2 \pi a_0^2$ for zero-energy Ps-He cross section.  The
22-pseudo-Ps-state calculation by McAlinden et al. yielded 13.193$\pi
a_0^2$. These results are in good agreement with each other. However, in a
recent study Ghosh and collaborators \cite{9a,9b} have argued the
importance of
including excited He states in a CC calculation. By including
a couple of excited states of He, they obtained a substantial reduction in
the zero-energy Ps-He cross section to 7.40 $\pi a_0^2$ in good agreement
with a model potential calculation by Drachman and Houston \cite{16} which
yielded $7.73\pi a_0^2$.  The coupled-channel model-potential calculation
by this author \cite{19} yielded $3.34\pi a_0^2$ calculated with the parameter
$C=0.84$
for the zero-energy Ps-He
cross section in agreement with a variational result (4.0 $\pm$
0.8)$\pi a_0^2$ \cite{ska2} and in total disagreement with another 
$10.4\pi a_0^2$ \cite{mt}.

Now we present a discussion of the results for Ps-He total cross sections
of the CC \cite{10,9a,9b}, $R$-matrix \cite{6} and model-potential
\cite{18,19} approaches shown in figure 1.  In the CC and model-potential
approaches the Born cross sections for Ps ionization and higher excitation
of the atom(s) are added to the result of the solution of the dynamical
equation. In the 22-pseudo-state $R$-matrix approach the Ps excitation and
ionization cross sections are obtained from the solution of the dynamical
equation. In figure 1 we also plot the different experimental cross
sections \cite{1,3,4,4a,4aa} as well as different zero-energy theoretical
\cite{dh,ska2,mt} cross sections. Of the experimental cross sections, the
total cross sections of Garner et al \cite{1,2a} and Skalsey et al
\cite{3} can be accommodated in a smooth graph. Once that is done it is
difficult to accommodate other experimental results in the same graph. The
model-potential cross section \cite{18,19} is the only one which is in
agreement with the results of Garner et al \cite{1,2a} and Skalsey et al
\cite{3} in the energy range $0 - 50$ eV. The 22-pseudo-state $R$-matrix
cross section \cite{6} is in agreement with the cross section of Nagashima
et al \cite{4} at low energies. The CC cross section without He states
\cite{10} is also in agreement with the cross section of Nagashima et al
at low energies. However, the CC cross section after including few He
states
\cite{9a,9b} is significantly reduced and is in agreement with the
theoretical model calculation of Ref. \cite{dh} and experimental cross
sections of Coleman et al \cite{4aa} and Canter et al \cite{4a} at low
energies.  Of the CC \cite{10,9a,9b}, $R$-matrix \cite{6} and
model-potential \cite{18,19} total cross sections the $R$-matrix result is
unique in not possessing a peak beyond the inelastic thresholds. The
combined experimental results of Refs. {1,3} exhibit such a peak in the
total cross section. To resolve the confusion in the low-energy cross
section of we consider below the pickoff quenching rate in Ps-He
scattering.

As the effective interaction for elastic scattering between Ps and He is
repulsive in nature, a smaller scattering length as obtained in in Refs. 
\cite{18,19} would imply a weaker effective Ps-He interaction. The scattering
length of the $R$-matrix \cite{6} and CC approaches without He states
\cite{10} is
1.90 a.u., model-potential coupled-channel approach \cite{19} is 0.91 a.u., 
the CC approach with few He states \cite{9a,9b} is 1.36 a.u., and a
model  calculation by Houston and Drachman \cite{16} is 1.39 a.u. A
small
scattering length as in the model-potential coupled-channel approach \cite{19}
would imply a weaker ortho Ps-He interaction and consequently, would allow the
ortho Ps atom to come closer to He which would lead \cite{22} to a large
pickoff quenching rate and a large $^1Z_{\mbox{eff}}$ $(\sim 0.11)$ in
agreement with experiment \cite{hy,4a,11a,11b,11c} as shown in figure 2,
where we plot
$^1Z_{\mbox{eff}}$ of different theoretical approaches and compare with
experiment. The static-exchange model leads to large low-energy cross sections
and hence a much too small $^1Z_{\mbox{eff}}$ \cite{12,13,14,22}.  The CC
\cite{10} and $R$-matrix \cite{6} models also yield a much too large scattering
length corresponding to a stronger repulsion between Ps and He.  Consequently,
these models should lead to a much too small $^1Z_{\mbox{eff}}$ $(\sim 0.04)$
in disagreement with experiment \cite{hy,4a,11a,11b}. This is addressed in
detail
in Ref.  \cite{22} where a correlation between the different
scattering lengths and the corresponding $^1Z_{\mbox{eff}}$ is
established. This correlation
suggests that a small Ps-He scattering length as obtained in the 
model-potential \cite{18,19} and variational \cite{ska2} approaches is
consistent with the large experimental $^1Z_{\mbox{eff}}$. 

\subsection{Ps-H$_2$ System}

The effective Ps-H$_2$ interaction is also repulsive in nature and there are no
known resonances in this system. There are two theoretical calculations of
cross section: one using the Born-Oppenheimer approximation by
Comi et al \cite{comi}, and the other using coupled-channel calculation with
model potential \cite{21,21a}. The two experimental results for total cross
section in this case are due to Garner et al \cite{1,2a} and Skalsey et al
\cite{3} which can be combined in a smooth curve, as can be found from figure
3. The low-energy cross sections of Comi et al \cite{comi} are orders of
magnitude larger than experiment. The theoretical total cross sections of Refs. 
\cite{21,21a} agree well with experiment in the low- to medium-energy region
for energies less than 20 eV.  In addition to the Ps(1s,2s,2p) excitations
included in the coupled-channel approach \cite{21,21a}, ionization and higher
excitations of Ps and some excitations of H$_2$ (B$^1 \Sigma_u^+$ and b$^3
\Sigma_u^+$ states)  \cite{21b} are included in Refs. \cite{21,21a} using
the
first Born approximation.

\subsection{Ps-Ne and Ps-Ar Systems}

The only dynamical calculation including exchange interaction in these systems
is the static-exchange calculation using the model potential \cite{20}. The
relevant experimental cross sections  are reported by Zafar et al
\cite{2b}, Skalsey et al
\cite{3}, and Coleman et al \cite{4aa}. In this case the theoretical scattering
length for Ps-Ar is 1.65 a.u., and for Ps-Ne is 1.41 a.u. The approximate
experimental scattering length obtained by Coleman et al \cite{4aa} for both
systems is 1.5 a.u. in reasonable agreement with theory.  The theoretical
cross
sections at low energies (0 to 5 eV) are in good agreement with the low-energy
experiment of Skalsey et al \cite{3}. 

\subsection{Ps-alkali-metal Atom Systems}

In Ps-alkali-metal atom systems so far there are no experiments. The CC method
has been applied in the static-exchange \cite{h1} and two-Li-state models
\cite{h2} for the Ps-Li system. There has also been a CC calculation by
the
Calcutta group reported at this workshop \cite{cal}. 
The coupled-channel approach with model
exchange potential using Ps(1s,2s,2p) states has been applied successfully for
Ps-Li \cite{pkb}, Ps-Na, Ps-K \cite{ska3}, Ps-Rb, and Ps-Cs \cite{ska4}
scattering. The CC calculation of Refs. \cite{h1,h2} did not report
results for
PsLi binding and resonances. The most interesting aspects of the model
potential calculation is the prediction for resonance and binding
energies
in
all the Ps-alkali-metal atom systems as well as the cross sections. In all
cases resonances were
reported in
singlet S, P and D waves near the lowest inelastic threshold. The
 binding
energies of these model calculations  are in agreement with other
theoretical calculations \cite{m1,m2}. 
 As no
other calculations or experiments are available for Ps scattering by these
systems, we do not
present a discussion of these results here and refer the interested readers to
appropriate places. 

\subsection{Ps-Ps system}

The model-potential
\cite{mp1} 
and the CC approaches \cite{cc1}  have been used in the three-Ps-state
model
for this
system. There has also been elastic scattering results using a variational
method \cite{mt}. In the overall singlet state the calculated scattering
lengths are $7.46a_0$ \cite{mp1}, 5.85$a_0$ \cite{cc1}, and 8.4$a_0$
\cite{mt} and those for spin-polarized ortho Ps ortho Ps  system
are 1.56$a_0$, $-49.43a_0$, and 2.95$a_0$, respectively. The results of
Refs. \cite{mt,mp1} are in reasonable agreement with each other. The
calculation of Ref. \cite{cc1} has produced a scattering length of wrong
sign corresponding to an attraction between two spin-polarized ortho Ps
atoms. The model potential calculation is unique in predicting resonances
in the overall singlet channel in S and P waves at
3.35 eV and 5.05 eV of widths 0.02 eV and 0.04 eV, respectively
\cite{mp1}. Both the model-potential and the variational methods yield the
correct result for Ps-Ps binding \cite{be8,mt,mp1}: 0.44 eV. The CC
calculation
underbinds the Ps-Ps system by a factor of 2 and leads to a Ps-Ps binding of
0.22 eV.

\section{Summary and Future Perspective}

The experiments of Ps scattering by different atoms and molecules
\cite{1,2a,2b,3,4,4a,4aa,4ab} have initiated a new era of theoretical
calculations of Ps scattering using $R$-matrix \cite{6,7}, CC
\cite{8a,8b,8c,8d,9,10}
and model exchange
potential \cite{17a,17b,18,19,20,21}
approaches in a coupled-channel framework. Of these, the model potential
approach has been the most successful in a proper description of
scattering in leading to total cross sections of Ps scattering, specially
at low energies, in different systems (Ps-H, Ps-He, Ps-Ne, Ps-Ar, and
Ps-H$_2$) in better agreement with experiment and variational calculations
compared to the two other approaches. This approach  has also  led to a
better description of binding and resonance energies in Ps-H, and
Ps-alkali-metal atom systems.

Much remains to be done in both theoretical and experimental fronts.
More exact calculations of low-energy elastic scattering are welcome. 
Experimentally, there are only results of total cross sections.
The data for partial, differential, and ionization  cross sections of  Ps-He,
and Ps-H$_2$
would help in the comparison of different calculations and will clearly
reveal which of the theoretical approaches are providing a better description
of Ps scattering. It would be better to concentrate at
low energies below the lowest inelastic threshold,  specially for Ps-He,
Ps-H$_2$, Ps-Ne, Ps-Ar, and different Ps-alkali-metal atom
systems. Precise
measurements  of Ps-He and
Ps-H$_2$ elastic cross sections would provide a  critical
test for theoretical approaches. 
It would be also interesting to verify if the prominent
resonances in Ps-alkali-metal atom systems  in Refs. \cite{ska3,ska4} can be 
observed experimentally from a measurement of cross section at low energies.
Finally, one should be prepared to undertake 
the challenging theoretical study of pickoff quenching rates of ortho Ps
interaction
with different targets. The experimentalists are far ahead in this topic
\cite{4ab,hy}.
There is
enough homework to be done  till the next workshop.

\vskip 5cm
Figure caption:

1. Total cross section of Ps-He scattering: data from Ref. \cite{1}
$\Box$, from Ref. \cite{3} $\bullet$, from Ref. \cite{4a} $\triangle$,
from Ref. \cite{4aa} $\ast$, from Ref. \cite{4} $\times$; calculation of  
Ref. \cite{16} +, of Ref. \cite{ska2} + with error bar, of Ref. \cite{mt}
$\Diamond$, 
of Ref. \cite{19}
full line, of Ref. \cite{6} dashed dotted line, of Ref. \cite{10} dashed
line, of Refs. \cite{9b} dashed-double-dotted line.

2. $^1$Z$_{\mbox{eff}}$ for Ps-He: data from Ref. \cite{4a} $\Box$, from Ref. 
\cite{hy} $\Diamond$,  
from Refs. \cite{11a,11b} +, from Ref. \cite{11c} $\times$, 
calculation of Ref. \cite{22} full line, of Ref. \cite{12} dashed line,
of Ref. \cite{15} dashed-dotted line.

3. Total cross section of Ps-H$_2$ scattering: data from Ref. \cite{1}
$\Box$, from Ref. \cite{3} $\bullet$, calculation of Ref. 
\cite{21} with target excitation  full line, without target excitation 
dashed line.
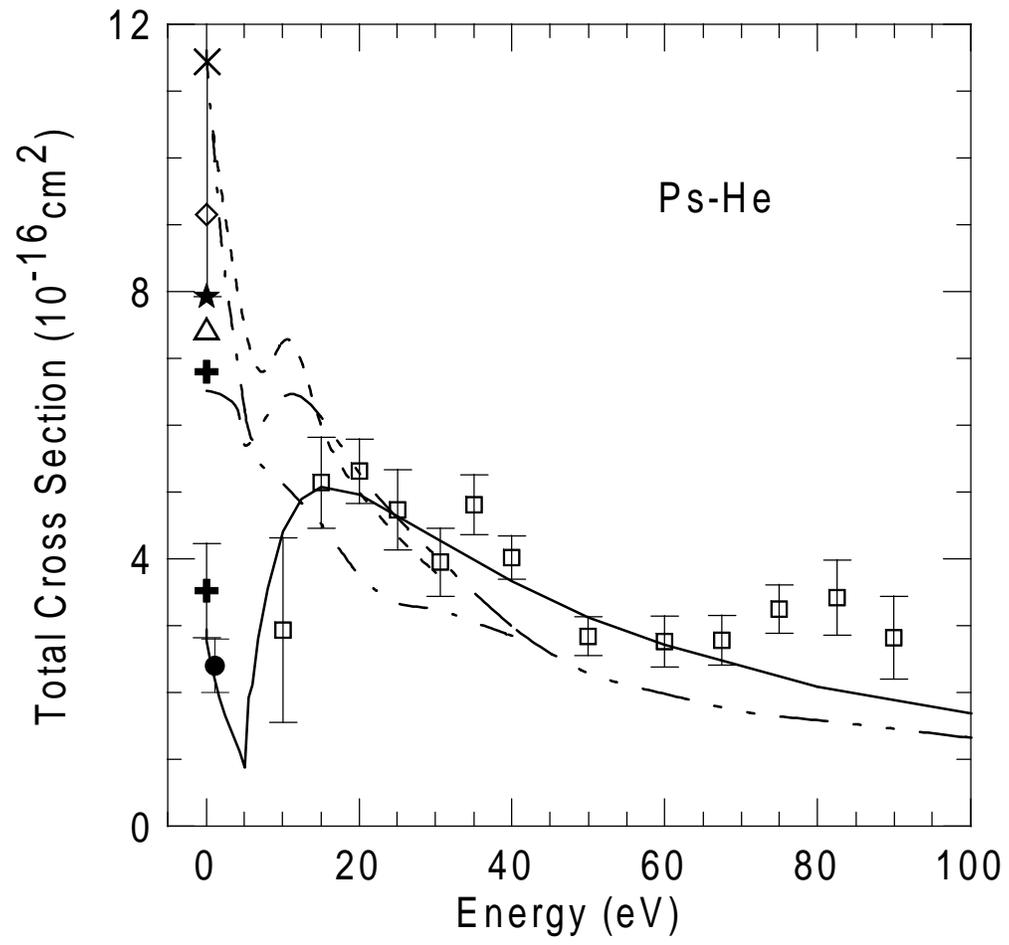
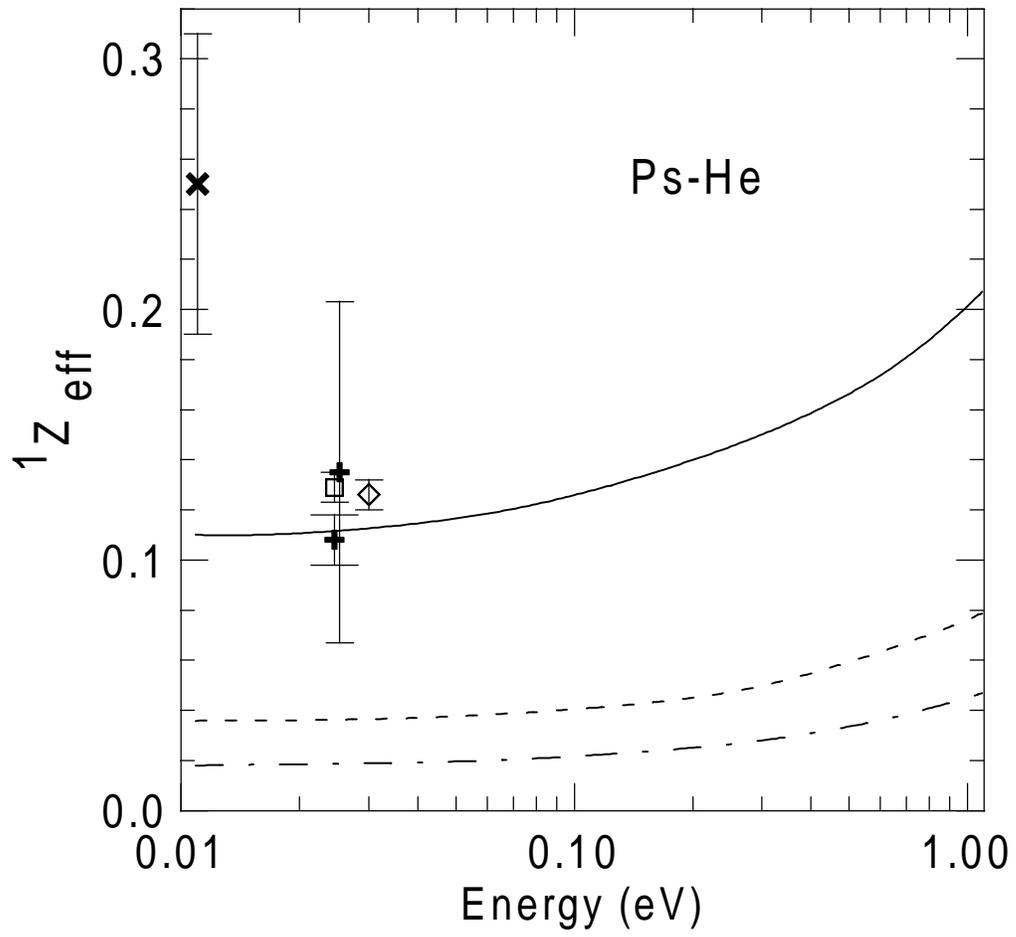
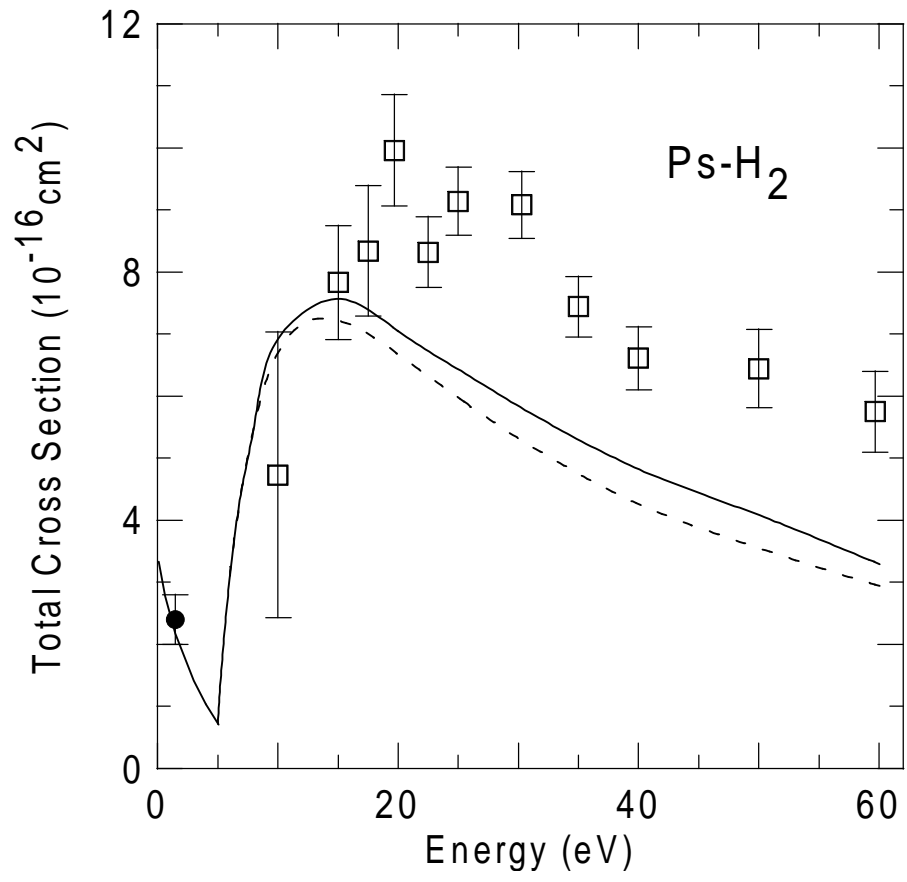
\end{document}